\documentclass[iop]{emulateapj-rtx4}
\usepackage{natbib}
\usepackage{color}
\usepackage{graphicx}
\newcommand{\Ha}{H$\alpha$}

\newcommand{\Msun}{$M_{\odot}$}

\shorttitle{X-Ray Expansion Velocity of the RCW\,86 Northeast Rim}
\shortauthors{Yamaguchi et al.}

\begin{document}

\title{The Refined Shock Velocity of the X-Ray Filaments in 
the RCW\,86 Northeast Rim}

\author{
Hiroya Yamaguchi\altaffilmark{1,2},
Satoru Katsuda\altaffilmark{3}, 
Daniel Castro\altaffilmark{1}, 
Brian J.\ Williams\altaffilmark{1},\\
Laura A.\ Lopez\altaffilmark{4},
Patrick O.\ Slane\altaffilmark{5},
Randall K.\ Smith\altaffilmark{5},
Robert Petre\altaffilmark{1}
}
\email{hiroya.yamaguchi@nasa.gov}

\altaffiltext{1}{NASA Goddard Space Flight Center, Code 662, Greenbelt, MD 20771, USA}
\altaffiltext{2}{Department of Astronomy, University of Maryland, College Park, MD 20742, USA}
\altaffiltext{3}{Institute of Space and Astronautical Science, JAXA, 3-1-1 Yoshinodai, Sagamihara, 
	Kanagawa 229-8510, Japan}
\altaffiltext{4}{Department of Astronomy and Center for Cosmology \& Astro-Particle Physics, 
	The Ohio State University, Columbus, OH 43210, USA}
\altaffiltext{5}{Harvard-Smithsonian Center for Astrophysics, 60 Garden Street, 
	Cambridge, MA 02138, USA}

\begin{abstract}
A precise measurement of shock velocities is crucial for constraining the mechanism and efficiency 
of cosmic-ray (CR) acceleration at supernova remnant (SNR) shock fronts. 
The northeastern rim of the SNR RCW\,86 is thought to be a particularly efficient 
CR acceleration site, owing to the recent result in which an extremely high shock velocity of 
$\sim$\,6000\,km\,s$^{-1}$ was claimed (Helder et al.\ 2009). 
Here we revisit the same SNR rim with the {\it Chandra} X-ray Observatory, 
11 years after the first observation. This longer baseline than previously available allows us to 
determine a more accurate proper motion of the nonthermal X-ray filament, revealing a much 
lower velocity of $3000 \pm 340$\,km\,s$^{-1}$  (and even slower at a brighter region). 
Although the value has dropped to a half of that from the previous X-ray measurement, 
it is still higher than the mean velocity of the \Ha\ filaments in this region ($\sim$\,1200\,km\,s$^{-1}$). 
This discrepancy implies that the filaments bright in nonthermal X-rays and \Ha\ emission  
trace different velocity components, and thus a CR pressure constrained by 
combining the X-ray kinematics and the \Ha\ spectroscopy can easily be overestimated. 
We also measure the proper motion of the thermal X-ray filament immediately to the south of 
the nonthermal one. The inferred velocity ($720 \pm 360$\,km\,s$^{-1}$) is significantly lower than 
that of the nonthermal filament, suggesting the presence of denser ambient material, possibly a wall 
formed by a wind from the progenitor, which has drastically slowed down the shock. 
\end{abstract}


%an average diameter of $\sim$$45'$. in various wavelengths (e.g., radio: Kesteven \& Caswell 1987, 
%X-rays: Pisarski et al.\ 1984, TeV $\gamma$-rays: Aharonian et al.\ 2009). 

\section{Introduction}

It is widely believed that supernova remnants (SNRs) are the predominant source of Galactic cosmic-rays (CRs) 
with energies up to $\sim$\,3\,PeV \citep[e.g.,][]{Reynolds08}. 
RCW\,86, probably the remnant of SN\,185 \citep[e.g.,][]{Clark75,Vink06}, is thought to be an efficient 
CR accelerator, based on the detection of nonthermal X-rays \citep{Bamba00,Borkowski01b,Rho02} and 
$\gamma$-rays \citep{Aharonian09,Yuan14}. 
The large amount ($\sim$\,1\,\Msun) of Fe ejecta in a low ionization state \citep{Yamaguchi11,Yamaguchi14b} 
and Balmer-dominated shocks \citep{Smith97} observed in RCW\,86 suggest a Type Ia supernova origin. 
The remnant is likely to have evolved in a wind-blown cavity formed by the progenitor binary system 
\citep{Williams11}, maintaining a high enough shock velocity to accelerate CR particles over the entire lifetime of the remnant.
In particular, its northeast (NE) rim is still expanding within the low-density cavity or has encountered 
the cavity wall very recently, so its current shock speed is the highest within the entire remnant 
\citep{Vink97,Yamaguchi08a,Broersen14}. For this reason, the NE rim of RCW\,86 has been frequently 
studied to understand the details of the CR acceleration mechanism.

A velocity of the SNR shock wave in combination with the physical condition of the postshock plasma offers 
a key for constraining the CR acceleration efficiency. The most direct way to determine the shock velocity 
is to measure the proper motion of the shock front. 
Using two {\it Chandra} observations of the RCW\,86 NE rim taken in 2004 and 2007, \cite{Helder09} 
(hereafter H09) obtained the proper motion of the nonthermal X-ray filaments over a three-year baseline to be $1''\!.5 \pm 0''\!.5$.  
This proper motion corresponds to a shock velocity of $6000 \pm 2000$\,km\,s$^{-1}$ at the well-agreed distance of 
2.5\,kpc \citep{Rosado96,Sollerman03}. H09 also analyzed an \Ha\ spectrum from the same region and 
derived a postshock proton temperature of $2.3 \pm 0.3$\,keV based on the width of the emission lines. 
This was unexpected because the theoretical relationship between the postshock temperature and shock velocity, 
$kT = (3/16) \mu m_p V_s^2$ (where $\mu$ and $m_p$ are the mean particle mass and the proton mass), 
predicts a much higher postshock temperature ($kT$ = 42--70\,keV depending on the degree of thermal 
equilibration) for $V_s$ = 6000\,km\,s$^{-1}$.
H09 attributed this discrepancy to the efficiency of the CR acceleration: a substantial fraction of the kinetic energy 
is being transferred into nonthermal energy. They concluded that $>$50\% of the postshock pressure 
is produced by the CR protons.

In their follow-up \Ha\ observation, however, the mean proper motion of the forward shock filaments was 
inferred to be $0''\!.10 \pm 0''\!.02$\,yr$^{-1}$, corresponding to a velocity of only $1200 \pm 200$\,km\,s$^{-1}$ 
\citep[][hereafter H13]{Helder13}. This velocity is far lower than their previous X-ray measurement, 
and consistent with the measured proton temperature ($\sim$2\,keV) with no energy injection to CR acceleration.  
H13 suggested that the \Ha-bright filaments are biased toward denser regions with decelerated shock velocities, 
while the X-ray-emitting filaments represent the relatively undecelerated regions of the shock. 
This scenario seems reasonable, given the fact that the forward shock of the SNR is interacting with 
an inhomogeneous medium \citep[e.g.,][]{Williams11}. In fact, H13 found substantial variation 
in the velocity of the \Ha\ filaments (ranging from 300 to 3000\,\,km\,s$^{-1}$). 
We note, however, that there is also a substantial uncertainty in the X-ray velocity measurement of H09, 
due to the small separation in time (only 3\,yrs) between the two {\it Chandra} observations. 
The positional shift they obtained ($1''\!.5 \pm 0''\!.5$) corresponds to only 3 $\pm$ 1~pixels of 
the {\it Chandra}/ACIS detector. 
An additional X-ray observation using a much longer baseline is therefore crucial for determining the precise shock velocity.

Here we present the latest {\it Chandra} observation of the RCW\,86 NE rim obtained in 2015, 
which is 11 years after the first observation.  
It reveals a much slower velocity of the nonthermal X-ray filaments than measured by H09. 
In \S2, we describe the observations and data analysis. We discuss the results in \S3, and present 
our conclusions in \S4. Since the aim of the present work is to report the accurate velocity of 
the forward shock, we focus exclusively on the imaging analysis in this {\it Letter}. 
Detailed spectroscopy and further studies will be presented in a future paper.
The uncertainties quoted in the text and the error bars in the figures represent 
the 1$\sigma$ confidence level, unless otherwise stated.

\section{Observations and Results}

We observed the NE rim of RCW\,86 on 2015 June 26 using the {\it Chandra}/ACIS-S chips. 
As summarized in Table\,\ref{table}, the aim point and roll angle were set almost identical to those 
in the first observation of this region in 2004 \citep{Vink06}, so that we can measure the proper motion 
as accurately as possible. We used CIAO version 4.8 and the latest calibration database (CALDB) for 
the data analysis. Examination of the light curves revealed no significant background flares. 
The net exposure we obtained was 67\,ks.

\begin{table}[t]
\begin{center}
\caption{Summary of the Observations.
  \label{table}}
  \begin{tabular}{lcc}
\hline \hline
Obs.\,ID & 4611 & 16952 \\
Observation Date & 2004 June 15 & 2015 June 26 \\
Instrument & ACIS-S & ACIS-S \\
Pointing R.A.\ (deg) & 221.27 & 221.28 \\
Pointing Dec.\ (deg) & --62.35 & --62.35  \\
Roll Angle (deg)  &  295.16 & 293.58  \\
Exposure Time (ks) & 71.7 & 67.2 \\ 
\hline
\end{tabular}
%\tablecomments{
%}
%\vspace{-3mm}
\end{center}
\end{table}

\begin{figure}[t]
  \begin{center}
	\vspace{1mm}
	\includegraphics[width=8.2cm]{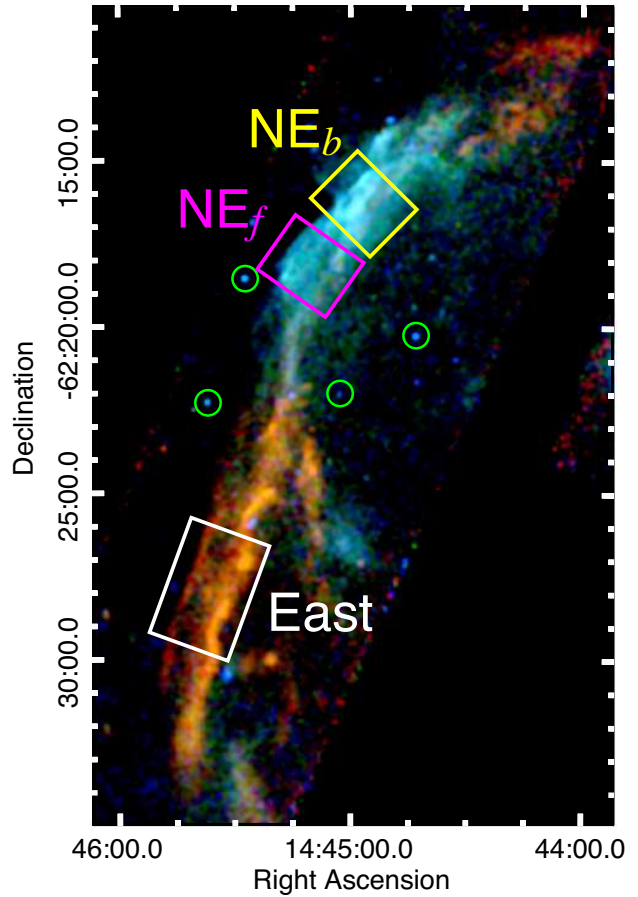}	
	%\vspace{1mm}
\caption{
Three-color X-ray image of the RCW\,86 NE rim obtained from the {\it Chandra}/ACIS-S 
observation in 2015. Red, green, and blue correspond to the energy bands of 
0.5--1.0\,keV, 1.0--2.0\,keV, and 2.0--5.0\,keV, respectively. 
The three rectangles indicate where we measured the proper motions. 
The background point sources indicated by the green circles are used for the coordinate realignment. 
  \label{image}}
  \end{center}
\end{figure}

\begin{figure*}[t]
  \begin{center}
	\vspace{1mm}
	\includegraphics[width=17.8cm]{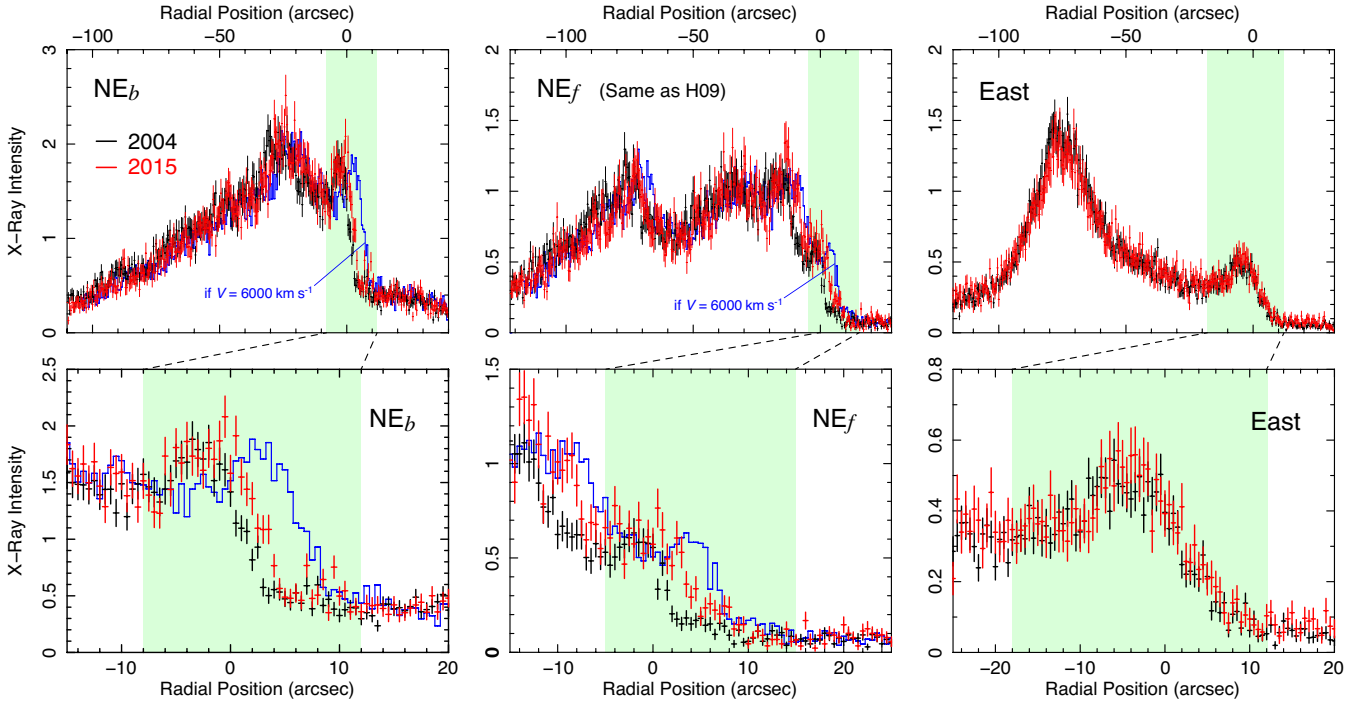}	
	\vspace{1mm}
\caption{
Projection profiles of the NE$_b$ ({\it left}), NE$_f$ ({\it middle}) and East ({\it right}) rims. 
Black and red indicate the data from the 2004 and 2015 observations, respectively. 
The unit of the vertical axis is $10^{-8}$\,photons\,cm$^{-2}$\,s$^{-1}$\,pixel$^{-1}$.
The shock front in the 2004 data is used as the origin of the horizontal axis. 
The bottom panels are the profiles magnifying the regions around the shock front. 
The data in the green regions are used to determine the proper motion of each rim. 
The blue lines in the left and middle panels are the profiles expected if the shock velocity 
is 6000\,km\,s$^{-1}$ as claimed by H09, which is plotted by simply shifting the 2004 profile 
5.5\,arcsec toward the positive direction. 
  \label{profile}}
  \end{center}
\end{figure*}

Figure\,\ref{image} shows the exposure-corrected ACIS image of the 0.5--1.0\,keV (red), 1.0--2.0\,keV (green), 
and 2.0--5.0\,keV (blue) bands, indicating that the dominant X-ray emission component changes from thermal (red) 
to nonthermal (blue) along the filaments from the south to the north. The previous work has suggested 
that the southern part has already encountered the dense wall of the wind-blown cavity, whereas the northern part is 
still in the low-density cavity and thus maintains a high CR acceleration efficiency 
\citep[e.g.,][]{Yamaguchi08a}, which we confirm below. 
For accurate determination of the X-ray proper motion, we compared the images from the 2004 and 2015 
observations, and aligned the coordinates using the relative positions of the four background point sources 
indicated in Figure\,\ref{image}. The details of the alignment procedure are described in \cite{Katsuda09}. 
The RMS of the residuals for the source positions between the two epochs is only $0''\!.22$, 
which we take as a systematic uncertainty in our proper motions measurements described below.
%The necessary shifts were $\sim0''\!.16$ in both Right Ascension and Declination, 
%negligibly small compared with the 11-yr proper motions whose measurements are described below.

We extract projection profiles of the nonthermal filaments from the yellow and magenta rectangles in Figure\,\ref{image} 
(hereafter ``NE$_b$'' and ``NE$_f$'', respectively). The latter (fainter filament) is exactly where H09 measured 
the shock velocity. 
Figure\,\ref{profile} ({\it left} and {\it middle}) shows the profiles from both the 2004 (black) and 2015 (red) observations, 
using data in the 0.5--5.0\,keV band. The shock front has clearly shifted over the 11 years in both filaments. 
Using the method established by \cite{Katsuda08}, we determine the proper motions of the NE$_b$ and NE$_f$
filaments to be $0''\!.150 \pm 0''\!.020$\,yr$^{-1}$ and $0''\!.253 \pm 0''\!.029$\,yr$^{-1}$, respectively. 
The corresponding velocity is $1780 \pm 240$\,km\,s$^{-1}$ for the NE$_b$ and 
$3000 \pm 340$\,km\,s$^{-1}$ for the NE$_f$ (at the distance of 2.5\,kpc), significantly less than 
the measurement by H09. 
For comparison, Figure\,\ref{profile} ({\it left} and {\it middle}) also shows the ``expected'' profiles 
(blue solid lines) for the case of $V_s$ = 6000\,km\,s$^{-1}$ (H09). 
An uncertainty in the distance to the SNR, $2.5 \pm 0.5$\,kpc \citep{Rosado96,Sollerman03}, 
corresponds to that in the shock velocity of $\pm 600$\,km\,s$^{-1}$ for the NE$_f$ filament 
($\pm 20$\% of the mean value). 
In addition, there is a systematic uncertainty of $\sim$\,260\,km\,s$^{-1}$ inferred 
from the astrometric accuracy ($0''\!.22$ between the two epoch). 
We also estimate another systematic uncertainty in our measurements by slightly 
changing the angle of the profile extraction regions, and obtain differences 
only within $\pm 0''\!.02$\,yr$^{-1}$ or $\pm 240$\,km\,s$^{-1}$. 
To summarize, the velocity of the NE$_f$ filament never exceeds 4500\,km\,s$^{-1}$ even with 
the multiple uncertainty components, and that of the NE$_b$ filament must be even lower.

The large time separation between the two observations also allows us to examine 
if the thermal-dominated filaments have indeed decelerated due to the collision with the dense material. 
The right panels of Figure\,\ref{profile} show the profiles extracted from the white rectangle in 
Figure\,\ref{image} (hereafter ``East''). Unlike the NE filaments, we see only a little shift of the shock front. 
The proper motion of this region is measured to be $0''\!.061 \pm 0''\!.031$\,yr$^{-1}$, corresponding to 
$720 \pm 360$\,km\,s$^{-1}$. We establish that the East rim has a lower velocity.  
The value is comparable to the \Ha-measured forward shock velocity at the southwest rim of 
this SNR where the shock front is also interacting with dense material 
\citep[$\sim$\,800\,km\,s$^{-1}$:][]{Rosado96}.

\section{Discussion}

Our new observation of RCW\,86 has revealed that the actual proper motions of the nonthermal 
filaments (NE$_b$ and NE$_f$) are significantly slower than previously measured by H09. 
However, the updated shock velocities are still inconsistent with the mean velocity derived from 
the \Ha\ proper motion in this region ($\sim$\,1200\,km\,s$^{-1}$: H13). This implies that the NE rim 
of RCW\,86 has a range of shock velocities due to a small-scale density inhomogeneity, 
and the \Ha-bright filaments are indeed biased toward the denser regions as suggested by H13. 
In fact, we have revealed the significant difference between the velocities of the NE$_b$ and NE$_f$ 
filaments in our analysis. In the thermal-dominated (East) region, on the other hand, 
the X-ray and \Ha\ velocities are consistent with each other, indicating the presence of dense material 
with a larger spatial scale, probably a wall formed by the progenitor wind \citep[e.g.,][]{Williams11}, 
that had decelerated the overall shock propagation. 
Assuming pressure equilibrium and thus constant $n_0 V_s^2$ along the shell (where $n_0$ is 
an ambient density), we roughly estimate the density ratio between the East and NE$_f$ regions to be 
$n_{\rm East} / n_{{\rm NE}_f} \approx 17$.

Ours is not the first result casting doubt upon the extremely high velocity ($\sim$\,6000\,km\,s$^{-1}$) claimed by H09. 
\cite{Williams11} compared some observed characteristics of RCW\,86 with their hydrodynamical 
simulations of a Type Ia SNR evolving in a wind-blown bubble. Although most of the characteristics 
were successfully reproduced with reasonable initial conditions, only the high shock velocity 
at the NE rim required more complex conditions, i.e., a significant offset between the explosion center 
and the SNR geometrical center.
Recent {\it Fermi} observations of RCW\,86 disfavor a hadronic origin for the $\gamma$-ray emission, 
where the inferred CR pressure is much lower than the estimate by H09 \citep{Lemoine12,Yuan14}. 
These discrepancies have likely been resolved by our result. 
Notably, the refined velocity of the nonthermal X-ray filaments (1800--3000\,km\,s$^{-1}$) 
is consistent with an older estimate from the typical photon energy of the synchrotron X-rays, 
under the assumption that the acceleration and loss timescales for 
the relativistic electrons are the same \citep{Vink06} 
(see also \cite{Castro13} for other regions of this SNR). 
This implies that the electrons responsible for the observed synchrotron X-rays 
have achieved their maximum energy at the NE region of RCW\,86.

From the relationship of $kT = (3/16) \mu m_p V_s^2$, a postshock proton temperature 
at the NE$_f$ region is derived to be $\sim$\,11\,keV, assuming thermal equilibrium 
(or higher if the equilibrium has not been achieved). 
Despite the substantial drop from the H09 value (i.e., $kT$ = 42--70\,keV), 
this temperature is still higher than that directly measured from the \Ha\ spectrum ($\sim$\,2.3\,keV: H13). 
One might, therefore, consider that a significant amount of the shock energy is injected into CR acceleration. 
However, since the \Ha\ emission is likely to be enhanced at denser regions where the shock velocity is lower, 
it could be inappropriate to combine the \Ha\ and {\it nonthermal} X-ray measurements for studies of 
shock waves interacting with highly inhomogeneous environment, like the NE rim of RCW\,86. 
We should also note that a substantial fraction of the upstream kinetic energy can be 
transferred into downstream turbulence, and thus the CR pressure estimated by the postshock 
temperature could be overestimated when naively compared with the shock velocity \citep{Shimoda15}.

One open question is if {\it thermal} X-ray emission from the swept-up ambient medium 
coincides with the \Ha-bright filaments in terms of both kinematics and plasma properties. 
Since \Ha\ and thermal X-ray spectra constrain postshock proton and electron temperatures, respectively, 
they have often been combined to determine the efficiency of collisionless thermal equilibration 
between the protons and electrons \citep[e.g.,][]{Rakowski03,Helder11}. 
Although the proper motion measurements on the RCW\,86 East (this work) as well as  
the northwestern rim of SN\,1006 \citep{Winkler03,Katsuda13} suggest a good coincidence 
between the thermal X-ray and \Ha\ filaments, this should be verified with adiditional observations 
of \Ha-bright forward shocks in other SNRs, like Kepler's SNR \citep{Sankrit16} and the Cygnus Loop \citep{Medina14}. 
Future observations with {\it Hitomi (ASTRO-H)} will allow us to determine the ion temperature of the swept-up medium. 
Comparison between it and the \Ha-measured proton temperature will also be useful for 
understanding the relationship between the thermal X-ray and \Ha\ observables.

\section{Conclusions}

We have presented new, improved measurements of the X-ray proper motion in the RCW\,86 
northeastern rim using {\it Chandra} data taken in 2004 and 2015 (with an 11-year separation). 
The velocity of the nonthermal filaments is derived to be 1800--3000\,km\,s$^{-1}$ 
(at the distance of 2.5\,kpc), which is much lower than the previous measurement by H09, 
but still significantly higher than the velocity inferred from the \Ha\ proper motion (H13).  
This implies different origins of the filaments bright in nonthermal X-rays and \Ha\ emission, 
with the former having a higher velocity. 
Caution should be exercised before combining nonthermal X-ray kinematics and \Ha\ spectroscopy to study shock physics, at least in a complex environment such as this. 
We have also shown that the thermal filament immediately to the south of the nonthermal filament has a velocity of 
$720 \pm 360$\,km\,s$^{-1}$, significantly slower than the nonthermal filament. 
The consistency between the velocities measured with the thermal X-ray and \Ha\ emission  
suggests a common origin, but this should be verified with more observational work.

\acknowledgments
We thank Dr.\ Ryo Yamazaki for helpful discussion at Harvard-Smithsonian Center for Astrophysics. 
This work is supported by the {\it Chandra} GO Program grant GO5-16072A.

\bigskip

%\bibliography{ms}

\end{document}